# Amplifying solid-state high harmonic generations with momentum k-gaps in band structure engineering

Yiming Pan[1], Danni Chen[1], Xiaoxi Xu[2], Zhaopin Chen[3,4,5], Huaiqiang Wang[6]

1. School of Physical Science and Technology and Center for Transformative Science, ShanghaiTech University, Shanghai 200031, China
2. School of Electrical Engineering—Physical Electronics, Center of Light-Matter Interaction, Tel Aviv University, Ramat Aviv 69978, Israel
3. Department of Physics, Technion – Israel Institute of Technology, Haifa 3200003, Israel
4. Solid State Institute, Technion – Israel Institute of Technology, Haifa 3200003, Israel
5. The Helen Diller Quantum Center, Technion – Israel Institute of Technology, Haifa 3200003, Israel
6. Center for Quantum Transport and Thermal Energy Science, School of Physics and Technology, Nanjing Normal University, Nanjing 210023, China

**Abstract**

We propose a novel amplification mechanism for high harmonic generation (HHG) in solids by leveraging bandgap engineering with momentum k-gaps. By constructing a simple diatomic lattice featuring balanced, alternating gain and loss profiles, facilitated by an array of four-level systems, we explore the physics of k-gap-amplified Bloch oscillations in the intraband channel of solid-state HHG. Through numerical simulations, we elucidate the coexistence of amplification and harmonic radiation processes in a solid. Our finding reveals that advanced bandgap engineering can define k-space optical devices - such as Brillouin cavity, Bloch-Zener oscillator and k-gap amplifier - thereby enabling the coherent manipulation of semiconductor radiation and high harmonic generation in both semiconductor superlattices and artificial materials. Furthermore, we analyze the spectrogram and material realizations required for amplifying solid-state HHG. These results underscore the potential of k-gap band structure engineering to advance coherent light sources at extremely short wavelengths.

The interaction of an intense electric field with matter – whether in atomic gases [1], liquids [2], or solids [3] - accelerates electrons within the medium, producing coherent radiations at extreme ultraviolet (EUV) and X-ray regimes. This process, termed high harmonic generation (HHG), enables direct observation and control of coherent electron dynamics, ushering in transformative advances in ultrafast quantum physics [4]. To enable high-precision pump-probe measurements of ultrafast electron dynamics [5,6], achieving high repetition rates and photon flux remains a central challenge in ultrashort pulse generation. However, conventional HHG processes exhibit extremely low efficiencies, typically on the order of $10^{-5}$ relative to the pump intensity [7]. Fundamental limitations persist for EUV/X-ray sources, including phase-matching conditions [8,9], unresolved amplification mechanisms for ultra-short wavelengths [10], and inherent inefficiency of methods such as superradiance [11]. Moreover, reliance on large-scale facilities like free-electron lasers and synchronization radiation [12,13] presents prohibitive costs and scalability barriers for compact ultra-short-wavelength light sources.

While the amplification mechanism for HHG in solids remains poorly understood, most efforts focus on overcoming technical challenges in EUV laser design. In contrast, we propose a paradigm shift: leveraging a small crystal to generate intense, coherent EUV light. Solid-state HHG, first experimentally realized in 2011 [3], provides distinct advantages over noble gas HHG, such as simpler fabrication, scalable integration, and reduced peak intensity requirements, attributable to the high electron density of solids. However, after a decade of material exploration – spanning semiconductors to dielectrics [14], the lack of an effective amplification scheme has confined solid-state HHG to low-efficiency regimes, limiting its utility as a viable EUV source.

In essence, solid-state HHG arises from two principal radiative channels: intraband currents [14], governed by field-driven acceleration of charge carriers within a single electronic band (e.g., super-Bloch oscillations), and interband polarization [15], in which strong-field tunneling generates electron-hole pair across different bands. Subsequent acceleration and radiative recombination of these pairs – mirroring the "three-step model" of atomic HHG – yield coherent emission. The dominant contribution between these channels is determined by the pump laser intensity and material band structures [16,17].

In this work, we present an unexpected mechanism to amplify solid-state HHG through momentum bandgap (k-gap) engineering. The concept of k-gaps – analogous to energy gaps ($\omega$-gaps) but characterized by "forbidden momentum regions" in the Brillouin zone – is inspired by recent advances in time-varying photonics [18,19] . Modes within k-gaps exhibit exponential growth over time, a phenomenon termed k-gap amplification [20]. While extensively studied in time-varying medium [20], including amplified emission and lasing, k-gap engineering remains largely uncharted in other material systems. Notably, k-gaps often emerge alongside exceptional points (EPs) [21–23] . Here, however, our focus is not on EPs or broader non-Hermitian physics but on utilizing k-gap engineering to enhance HHG in solids. Our framework posits that Bloch band electrons, driven by the pump, traverse a k-gap and experience a population increase in their intraband current, leading to radiative amplification. This prompts a fundamental question for non-Hermitian electronic systems: what exactly undergoes amplification – the pumped electron or the radiated photon? To address whether this amplification manifests as an increase in carrier density or radiation intensity, we construct an effective tight-binding diatomic lattice that features balanced gain and loss profiles using a four-level recycling process (Fig. 2a). This construction ensures that the total carrier population is conserved, even as the radiative emission intensity of Bloch electrons under a driving force is amplified. We then investigate the gain profile and radiative spectrogram of solid-state HHG. Ultimately, this k-gap-amplified HHG paradigm offers a promising route toward compact, coherent extreme ultraviolet light sources.

**Bloch radiation of band electrons.** Bloch oscillation in crystalline solids can be interpreted as radiation emitted by an electron undergoing classical acceleration, similar to the radiation of an oscillating dipole (or current) [24–27]. This interpretation connects the Bloch theorem with classical radiation theory, positioning Bloch oscillations as a fundamental mechanism for photon emission in periodic potentials – a concept crucial to the subsequent analysis. In the presence of an external field, a band electron traversing the lattice accumulates energy, given by $\Delta E = eEa$, where $E$ is the applied field amplitude, $a$ is the lattice constant, and $e$ is the elementary charge. The Bloch theorem enforces periodicity, requiring that the electron releases this energy gain upon returning to its initial state. Energy dissipation occurs through various channels, including thermal coupling to lattice vibrations, non-radiative transfer to other electrons (e.g., Auger scattering), or coherent radiation at the Bloch frequency $\omega_B = eEa/\hbar$. Crucially, this radiation — termed Bloch radiation, provides a coherent mechanism for light emission, directly linking electron dynamics in solids to radiative processes [28].

The bandgap engineering for k-gap-amplified Bloch oscillations is as follows. The Bloch theorem dictates that band electrons oscillate within the Brillouin zone, acting as a resonator that selects specific radiative modes. When a DC field is applied, electrons move periodically in momentum space, emitting radiation coherently in the form of classical oscillating currents or dipoles. The Brillouin zone thus functions as an optical micro-ring ("k-space ring"), as depicted in Fig. 1b. Dimerization of the simple lattice splits a single band into two, reducing the Brillouin zone from (-π/a, π/a) to (-π/2a, π/2a), where the dimerized lattice constant is 2a. When band electrons, driven by a gradient field, cross the reduced Brillouin zone, they may undergo non-adiabatic splitting through Landau-Zener tunneling (LZT) [29–31], analogous to a beam splitter. The LZT probability depends on the ratio of the driving field strength to the energy gap induced by dimerization. Alternatively, by introducing an on-site gain/loss profile into the lattice, we can engineer a k-gapped band structure (Fig. 1d). When electrons traverse these k-gaps periodically in momentum space, their oscillations are amplified. In this sense, this k-gap engineering offers a promising pathway to facilitate the lasing of solid-state HHG through the amplification of Bloch radiation.

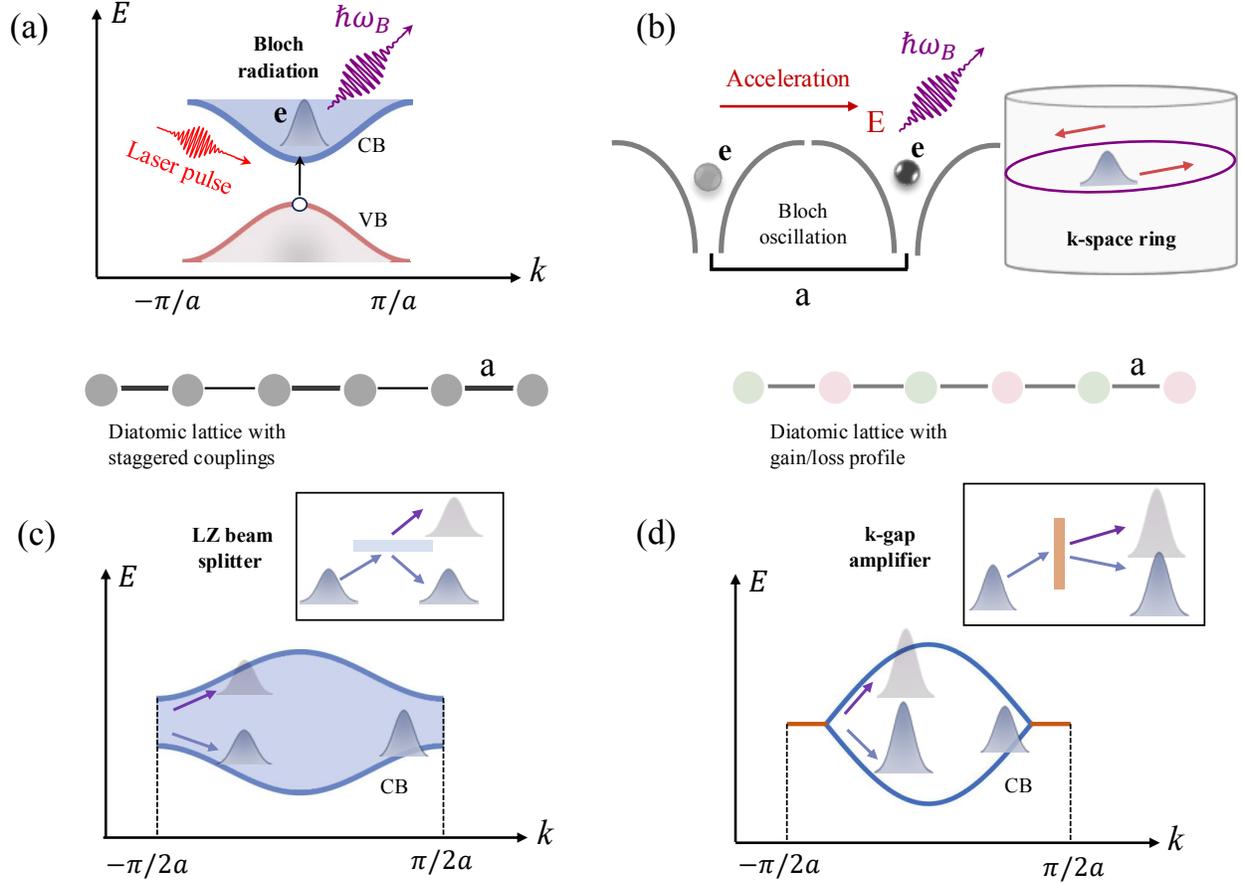

**Figure 1** Illustration of three optical devices constructed using Bloch oscillations in band structure engineering with designable energy gaps or momentum k-gaps. (a) an optical microring (Brillouin zone). (b) a beam splitter (Landau-Zener tunneling), and (c) a k-gap amplifier due to the travelling across the k-gap region.

**Modelling of k-gaps in solids.** To introduce the momentum k-gaps in an electronic lattice system, we begin by designing balanced gain and loss profiles for each site. The setup is analogous to an array of the atomic four-level system, where two short-lived virtual energy levels are introduced for the gained site (G-site) and the lossy site (L-site), as illustrated in Fig. 2a. For the G-site, we incorporate a virtual $N_3$-level. Electrons are pumped to this level by UV light, quickly relaxing to the G-site and accumulating population there. Conversely, for the L-site, we introduce a virtual $N_1$-level. Electrons are then pumped out of the L-site to the $N_1$-level, which has a short lifetime, enabling the population to relax back to a reservoir, denoted as the ground state $N_0$). This construction results in a similar four-level system, with the G- and L-site collectively forming the $N_2$-level as a diatomic site. Gain at the G-site arises from relaxation from $N_3$-level, while loss at the L-site is due to relaxation to $N_1$-level. Notably, the hopping between sites is described by a quantum-mechanical process, distinct from relaxation. By applying rate equations, we ascertain that the total population of all levels is conserved, expressed as $\frac{d}{dt}N_{tot} = 0$, where $N_{tot} = N_3 +$ $N_G + N_L + N_1 + N_0$. The local population of G-site $N_G(t)$ (and L-site $N_L(t)$) can exhibit exponentially growth or decay, controllable by a UV light pump (from $N_0$ to $N_3$ level). When these G/L sites are arranged in sequence, they form a one-dimensional diatomic lattice with balanced gain/loss profiles (Fig. 2b). Using tight-binding approximation, the Hamiltonian of this lattice is given by

$$H_0 = \sum_{j=1}^{N}\left(i\gamma c_{jG}^\dagger c_{jG} - i\gamma c_{jL}^\dagger c_{jL}\right) + \sum_{j=1}^{N-1} \kappa\left(c_{jG}^\dagger c_{jL} + c_{jL}^\dagger c_{j+1G}\right) + h.c., \quad (1)$$

where $\gamma$ represents the effective balanced gain and loss factor, and $\kappa$ is the coupling strength between the adjacent G- and L- sites. In momentum space, the Hamiltonian is $H_0 = \sum_k H_k C_k^\dagger C_k$, where the operator is $C_k = (c_{kG}, c_{kL})^T$, with $H_k$ being given by

$$H_k = \begin{bmatrix} -i\gamma & 2\kappa \cos(ka) \\ 2\kappa^* \cos(ka) & i\gamma \end{bmatrix}. \quad (2)$$

From this, we derive the energy dispersion $E_k = \pm\sqrt{4\kappa^2 \cos^2(ka) - \gamma^2}$. This k-gapped band structure is depicted in Fig. 1d. Exceptional points (EPs) arise at $E_k = 0$, given by the condition $\cos(k_{EP}a) = \pm\gamma/2\kappa$. Assuming a small gain/loss factor, $\gamma/2\kappa \ll 1$, the momentum positions of the EPs are $k_{EP} = \pm\left(\frac{\pi}{2a} - \frac{\gamma}{2\kappa a}\right)$. For $k \in \left(-\frac{\gamma}{2\kappa}, \frac{\gamma}{2\kappa}\right)$, the energy $E_k$ remains real, whereas for $k \in \left(-\frac{\gamma}{2a}, -\frac{\gamma}{2\kappa}\right) \cup \left(\frac{\gamma}{2\kappa}, \frac{\gamma}{2a}\right)$, the energy becomes imaginary, corresponding to the emergence of the k-gap. By considering the momentum periodicity of the Brillouin zone, the "forbidden" regions of k-gap are approximately $\Delta_k = \gamma/\kappa a$.

Notably, we examine the relationship between EPs and k-gaps. For instance, if in the four-level setting involving only a pair of G/L sites, the Hamiltonian reduces to

$$H_{G/L} = \begin{bmatrix} -i\gamma & \kappa \\ \kappa^* & i\gamma \end{bmatrix}, \quad (3)$$

which is structurally same to Eq. 4. The corresponding eigenvalues $E = \pm\sqrt{\kappa^2 - \gamma^2}$. EPs occur when $\gamma > |\kappa|$, and they can manifest in parameter spaces. In contrast, k-gaps are specific to momentum space. Such parity-time (PT) symmetric systems have been experimentally realized in coupled waveguides [32]. Notice that, EPs are not observed when $\gamma \ll |\kappa|$. Nevertheless, in the setting of our lattice construction, specific Bloch band modes that satisfy $|\cos(k_{EP}a)| < \gamma/2\kappa$ can fail into the k-gap region, and thus, being amplified.

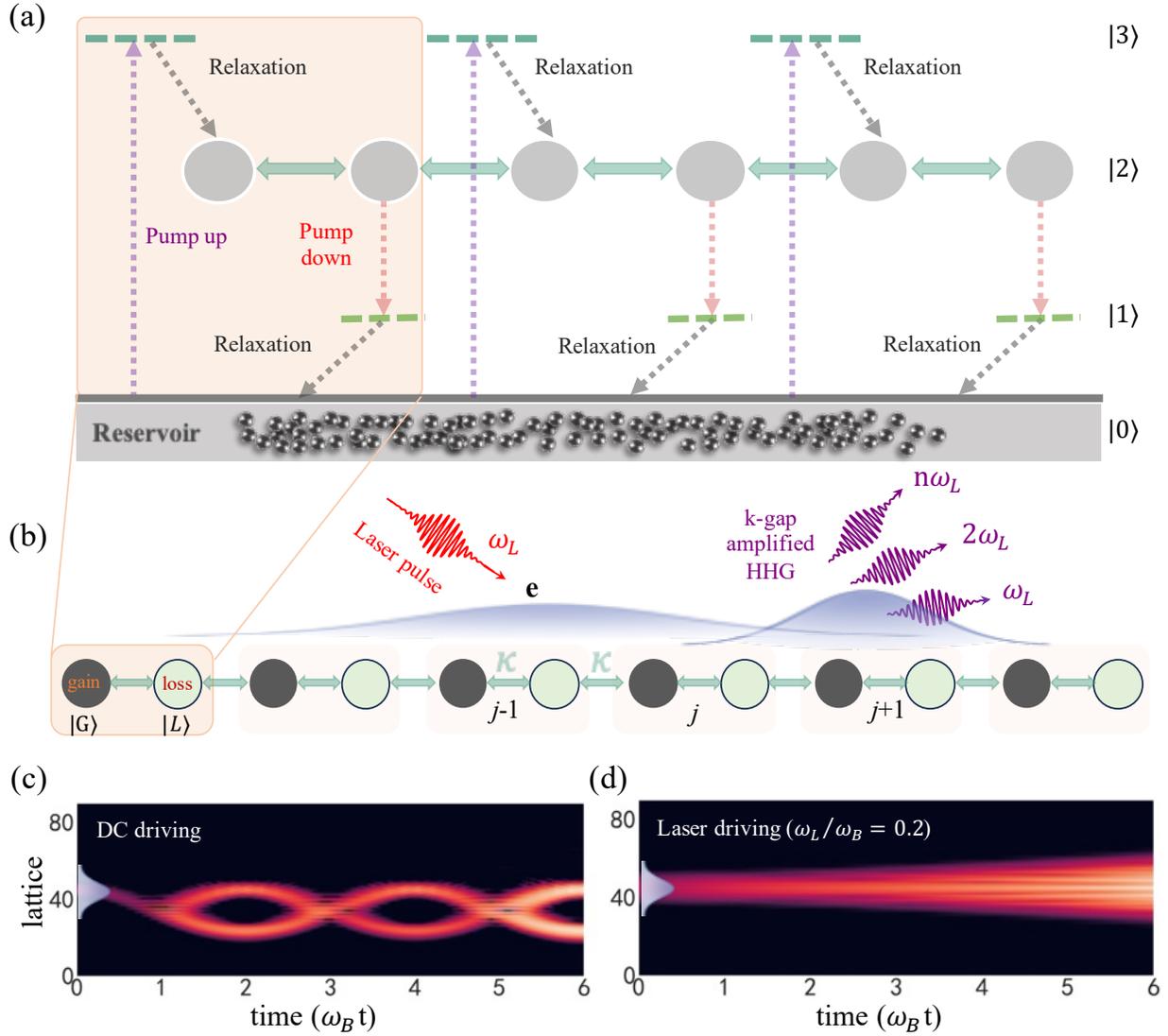

**Figure 2** Four-level construction of one-dimensional diatomic lattice with balanced gain/loss profiles and the dynamics of k-gap-amplified Bloch oscillations. (a) We construct the gain/loss profiles by introducing two auxiliary virtual levels ($N_1$ and $N_3$-levels). (b) The schematic setup of amplifying HHG in the gain/loss profiled diatomic lattice. (c) An amplified Bloch oscillation under a DC driving field. (d) An amplified super Bloch oscillation under an AC/laser driving field. This leads to the key realization of amplifying solid-state HHG.

**K-gap amplification.** To account for the effects of a laser field applied to the lattice, the Hamiltonian is modified as follows: $H = H_0 - F(t)\sum_{j=1}^{N}(2j)c_{jG}^{\dagger}c_{jG} + (2j+1)c_{jL}^{\dagger}c_{jL}$. In the context of semiclassical wavepacket evolution, the driving force leads to the equation of motion for the Bloch electron $\hbar\frac{dk}{dt} = F(t)$, where $F$ is the driving field. For a DC electric field ($F = -eE_0$), the time-dependent momentum is $k(t) = k(0) - eE_0 t/\hbar$, while for a laser field (or AC) ($F = -eE_0\cos\Omega t$), $k(t) = k_0 - eE_0\sin(\Omega t)/\hbar\Omega$, which can be obtained by the canonical substitution $p \to p - eA(t)$. Bloch oscillation of electrons can lead to Bloch radiation. We derive the group velocities via the dispersion: $v_g^{(\pm)} = \partial E_k/\partial k = \mp\frac{4\kappa^2 \cos ka \sin ka}{\sqrt{4\kappa^2\cos^2(ka)-\gamma^2}} = dx(t)/dt$. The sign denotes the upper band branch ($v_g^{(+)}$) and the lower band branch ($v_g^{(+)}$). From the last term, we can extract the classical trajectory of the wavepacket in real space, leading to the dipole moment $P(t) = -ex(t)$, which reflects the wave-particle duality of electron wavepackets [33]. The time-averaged radiation power emitted by a classical dipole is given by [34]: $I_P(\omega) = \frac{\mu_0\omega^4}{12\pi c}|\mathcal{F}[P(t)]|^2$, where $\mathcal{F}[P(t)]$ represent the Fourier component of the moment. Finally, we integrate group velocity over time and obtain the dipole moment:

$$P(t) = -e\int v_g(t')dt' = -\frac{\hbar}{2E_0}\left(\sqrt{4\kappa^2\cos^2(k(t)a)-\gamma^2} - \sqrt{4\kappa^2\cos^2(k(0)a)-\gamma^2}\right) \quad (4)$$

For $\gamma = 0$, the expression in Eq. 4 describes a conventional Bloch oscillation and the associated radiation power can be derived as $I_P(\omega) = \frac{\mu_0\omega_B^4}{12\pi c}\frac{\hbar^2\kappa^2}{E_0^2}$. However, for $\gamma \neq 0$, the dipole moment becomes imaginary as $k(t)$ approaches the exceptional points, leading to unphysical behavior. This occurs because the group velocity $\left|v_g^{(\pm)}\right| \to \infty$ diverges arounds the EPs. Despite this, the causality principle ensures that the electron cannot exceed the speed of light. The imaginary component of $\langle x(t)\rangle$ arises from the evolution of the wavepacket through the k-gap. This group velocity divergence signifies the complexity of Bloch oscillations within a k-gapped energy dispersion, suggesting that physical observables may become ambiguous due to the complex-valued dipole moment (Eq.4).

To understand why the semiclassical approach fails in this scenario, we numerically simulate an electron evolution in the real lattice. Figure 2c and 2d depict the wavepacket evolution under both DC and laser fields. These plots illustrate how the electron wavepacket moves through the k-gap, with its central trajectories in real space associated with the corresponding band structure. Far from the k-gap, the wavepacket oscillates in the lattice. However, as it enters the k-gap region, the wavepacket seems frozen and begins to grow exponentially in amplitude (see the Movie in the SM file). This exponential growth within the k-gap has been extensively discussed in photonic time crystals [35].

Still, we can properly interpret the process when adding the k-gap amplification in the semiclassical picture. When a Bloch electron is driven by an external force with all its k-components lying in the k-gap, the energy of band modes become purely imaginary, resulting in the global growth over time. Although both k-gap amplification and attenuation occur simultaneously in k-gap, the amplification would predominate in long term. Within the k-gap, the real part of the energy is fixed and the group velocity becomes zero ($v_g = \partial \Re(E_k)/\partial k = 0$). leading to the frozen phase while the amplitude continues to grow. Over time, all k-components exit the k-gap and return to the band. Upon crossing the EP, the amplified wavepackets splits into two: one component is scattered to the upper band, while the other is scattered into the lower band. As depicted in Fig. 2c, we can summary the process: upon entering the k-gap, the wavepacket pauses and undergoes amplification. As the amplified wavepacket, driven by an external force, leaves the k-gap and crosses the EP, it splits into two wavepackets, each residing on different band. These wavepackets oscillate on the bands, subsequently reach the EP again and re-enter the k-gap region for the next round of amplification. This cycle - comprising amplification, splitting, oscillation, and re-amplification – is referred to as amplified Bloch oscillation.

In the case of a laser-driven system, Figure 2d illustrates the k-gap-amplified solid-state HHG, also known as super Bloch oscillation in AC field driving. Although the evolution pattern appears chaotic, it adheres to the same cycling process as the amplified Bloch oscillation. However, in this case as the driving field alternates periodically, different k-components of the wavepacket enter the k-gap region at different times, so that the amplification and oscillation processes become more intertwined and less distinguishable. Therefore, the full wavepacket evolution looks messy without showing obviously trajectories, as compared to Fig. 2c.

**Quadrupole radiation.** When using semiclassical wavepacket dynamics to calculate the classical dipole moment, challenges arise when the displacement x(t) becomes imaginary. This issue necessitates a more appropriate method to calculate the radiation and solid-state HHG associated with the amplified Bloch oscillation. Numerically, we can define the dipole moment in lattice space as $P(t) = -e \int \psi^*(x,t) x \psi(x,t)\, dx = -ea \sum_j j c_j^\dagger(t) c_j(t)$. However, when this definition is applied to these simulation results, the dipole moment is found to be zero, as illustrated in Fig. 2c, 2d. This outcome results from the symmetric evolution pattern around its center. As the wavepacket splits evenly at the exceptional points, with one half moving left and the other moving right, their contributions to the dipole moment cancel each other out. In addition, even for conventional case ($\gamma = 0$), we see that a spatially localized, point-like electron input would exhibit Bloch breathing behavior, also evolving symmetrically, and as a result, the calculated dipole moment trivially yields P(t)=0. Therefore, even without k-gap engineering, direct use of the classical dipole radiation looks invalid.

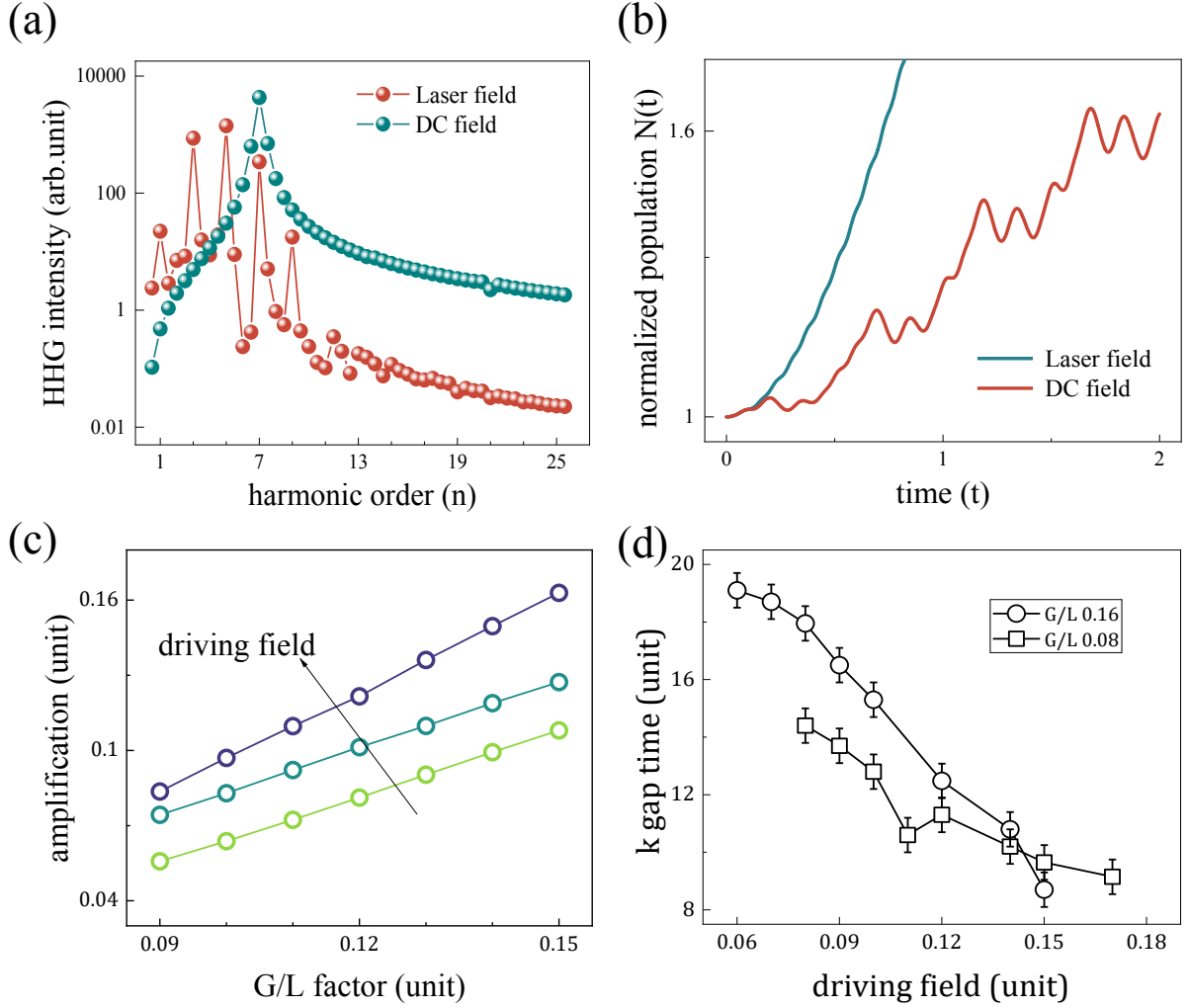

**Figure 3** Quadrupole radiation in amplified Bloch oscillation with k-gapped band structure engineering. (a) Comparison of AC- and DC-driven spectral characteristics. The cutoff frequency of high harmonic generation relates to the Bloch oscillation frequency. (b) The change of population N(t) over time. The flat segment indicates evolution on the band, while the gain segment shows dynamics in the k-gap. The slopes represent the effective gain coefficient Γ and the duration represents the k-gap time $\tau_{gap}$. (c) Linear relationship between the effective gain coefficient Γ and the gain/loss profile γ for three force strengths: F = 0.536, 0.766, 0.996 ×10⁻¹ eV/a (d) The k-gap time $\tau_{gap}$ inversely correlates with the driving field strength, matching theoretical expectations.

We observe that while the center of the wavepacket does not oscillate, its width does. Therefore, we define a radiation source based on this oscillating width, $Q(t) = -e\langle x^2 \rangle = -e \int \psi^*(x,t) x^2 \psi(x,t)\, dx = -ea^2 \sum_j j^2 c_j^\dagger(t) c_j(t)$. This is referred to as the quadrupole moment, corresponding to an oscillating linear electric quadrupole. Analogous to dipole radiation, we can directly obtain the time-averaged radiation power from a classical quadrupole [34][Griffiths' textbook]:

$$I_Q(\omega) = \frac{\mu_0 \omega^6}{60\pi c^3} |\mathcal{F}[Q(t)]|^2 \qquad (5)$$

To refine the definition of quadrupole radiation for amplified Bloch oscillations, we decompose the quadrupole moment into two parts: $Q(t) = \bar{Q}(t)N(t)$, where $\bar{Q}(t)$ is a normalized quadrupole and $N(t) = \int |\psi(x,t)|^2 dx$ is the population, so that $\bar{Q}(t) = Q(t)/N(t)$. The normalized quadrupole $\bar{Q}(t)$ is instrumental for analyzing the HHG spectrum, while the accumulated population $N(t)$ signifies the amplification. For each Bloch oscillation cycle, population growth can be approximated as $N(T_B)/N(0) = e^{\Gamma \tau_{gap}}$, where $\Gamma$ is the effective gain and $\tau_{gap}$ is the k-gap time. The effective gain and k-gap time depends on the balanced gain/loss factor, the k-gap size and the driving force strength. Notice that quadrupole radiation has been used in the analysis of electron dynamics and harmonic generations driven by bright squeezed vacuum [36].

Figure 3 compare quadrupole radiation from DC-driven Bloch oscillation and laser-driven HHG. In Fig. 3a, the HHG cutoff frequency is shown to scale linearly with the Bloch frequency $\omega_B$, increasing with the driving force $E_0$, and Fig. 3b depicts the temporal evolution of populations N(t) under both driving conditions. The effective gain $\Gamma$ is derived from the slope of N(t) in the gain region, with its duration corresponds to the k-gap time $t_{gap}$. For DC cases (Fig. 3c), $\Gamma$ exhibits a near-linear dependence on the gain/loss profile $\gamma$. Theoretically, $t_{gap} = \hbar \Delta_k/eE_o = \gamma/\kappa \omega_B$, inversely proportional to $\omega_B$ or $E_0$, a trend corroborated by numerical simulations (Fig. 3d). The simulation parameters see the SM file.

To resolve temporal radiation characteristic across the k-gap and energy bands, the Gabor transformation is applied to the averaged quadrupole moment: $I(\omega, \tau) = \left| \int \bar{Q}(t) W(t + \tau) e^{i\omega t} dt \right|^2$, where $W$ is a temporal window function. These resulting numerical spectrograms (Fig. 4a-4c), reveal HHG-like spectral broadening of frequency distributions with increasing driving force F=6.8931, 5.3613, and to 3.8295×10$^{-1}$ eV/a. Oscillatory intensity growth dominates in Fig. 4a and 4b, while Fig. 4c shows merely oscillation, reflecting competition between k-gap-induced population amplification and Bloch radiation. Fig. 4d-4f demonstrate intensity enhancement strongly correlated with the gain/loss profile $\gamma$, peaking under DC driving. Dipole and quadrupole radiation spectrum (Fig. 4g, 4h) align for both wide and narrow wavepackets, with characteristic frequencies consistent with inter-band HHG radiations, as compared in Fig. 4i. The spectral consistency validates the applicability of the quadrupole radiation in our modelling.

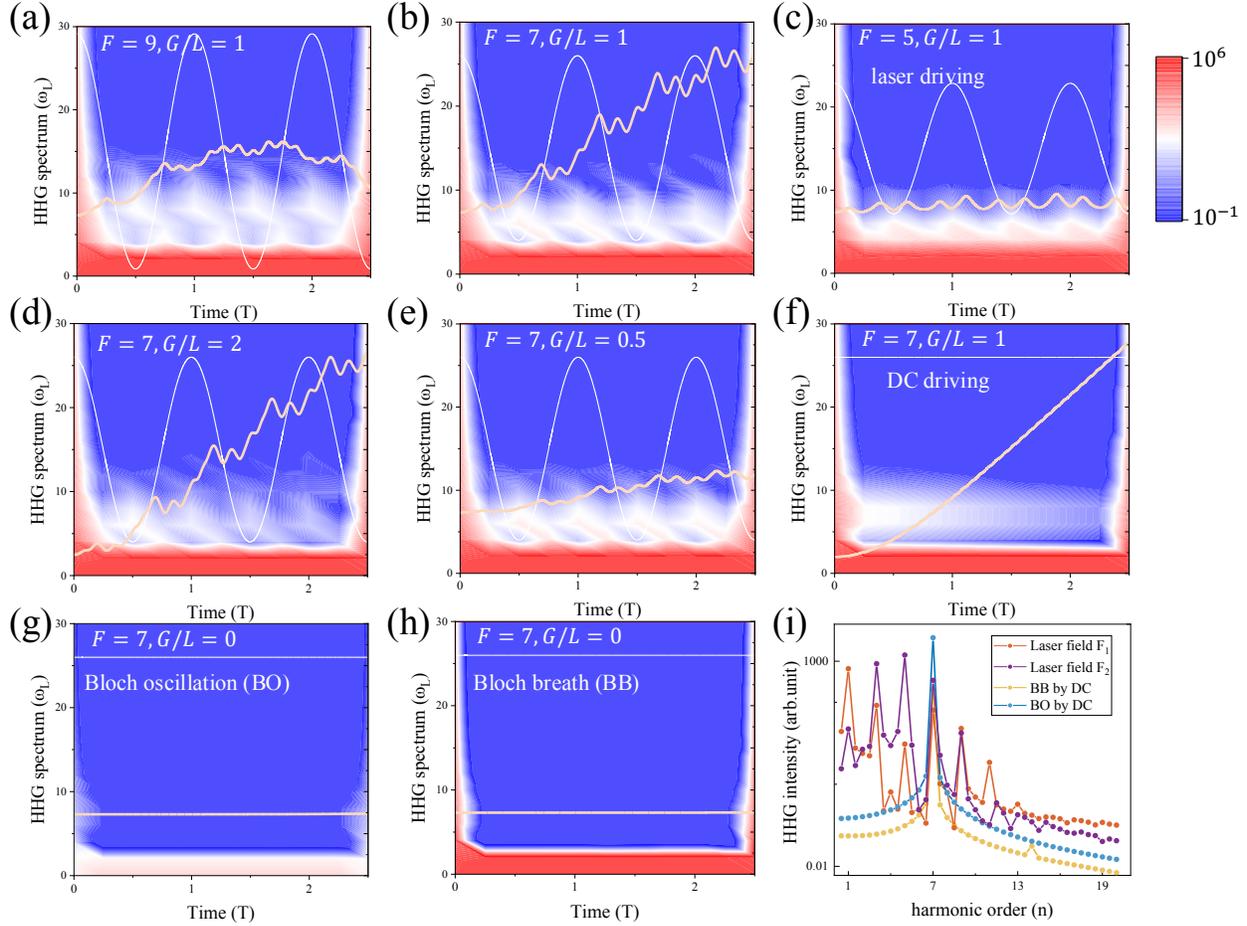

**Figure 4** Spectrogram analysis of the k-gap amplified HHG in solid. Different HHG spectrum are compared in terms of the driving field parameters and gain/loss profile. (a)-(c) HHG spectra under different driving field strengths (F). The white line shows F over time, the yellow line shows population dynamics, and the color map shows the short-time Fourier transform (STFT) spectrogram. The population dynamics combine k-gap amplification and HHG oscillation. (d)-(e) HHG spectra under different gain/loss profiles ($\gamma$), showing a strong link between $\gamma$ and amplification efficiency. (f) Spectrum under DC driving, showing higher enhancement compared to AC driving. (g)-(h) Comparison of dipole radiation (Bloch oscillation) and quadrupole radiation (Bloch breathing) without k-gaps. Both show similar spectral distributions, validating the quadrupole model for centrosymmetric electron dynamics. (i) HHG spectra under different F values (left) and comparison of Bloch oscillation (BO) and Bloch breathing (BB) radiation (right). The cutoff frequency and peak positions change with F. The consistency between quadrupole and dipole models further supports the quadrupole radiation mechanism. (F unit: $0.766\times10^{-1}$ eV/a; $\gamma$ unit: $0.244\times10^{-1}$ eV)

**Further discussions.** Several theoretical frameworks have been developed to model solid-state HHG [37], including the time-dependent Schrödinger equation (TDSE) [38] with a periodic potential, semiconductor Bloch equations (SBEs) [15], and time-dependent density functional theory (TDDFT) [39]. Although SBEs and TDDFT are particularly effective for studying specific materials, given their ability to accurately simulate the band structure via density functional theory, the tight-binding approximation (TBA) Hamiltonian approach, employed as a phenomenological model for solving the TDSE, provides a clearer understanding by capturing the essential k-gapped band structure engineering. Therefore, TBA model serves as an excellent candidate for discussing the HHG amplification in solids.

On the other hand, several experimental challenges must be addressed. First, the construction of the required four-level system to establish a balanced gain and loss profile may surpass the current capabilities of available materials. Potential platforms for such material design include aligned quantum wells or quantum dots with balanced gain/loss characteristics, optically pumped materials that exploit stimulated emission in doped semiconductor layers, and time-varying photonic crystals and metamaterials engineered to exhibit non-Hermitian responses. These systems would enable the controlled k-gap amplification of Bloch oscillations, facilitating the realization of efficient solid HHG. Second, maintaining electron coherence during the cycle of the amplified Bloch oscillation presents an additional challenge, particularly if the band electrons are subject to strong decoherence and dephasing. Third, generating coherent HHG in the extreme ultraviolet range, for instance, which requires Bloch frequencies between 10 to 100 eV, necessitates an electric field strength on the order of 1 GV/m and a superlattice constant ranging from 10 to 100 nm. Notably, Bloch oscillations have found applications such as THz sources in superlattice-based Bloch oscillators and quantum cascade laser (QCL) utilizes intrasubband transitions in quantum wells [28,40,41], commonly used in mid- to far-infrared electromagnetic radiation. Collectively, these stringent requirements push the limits of microfabrication technologies and material science, underscoring the complexity and technical hurdles in enhancing solid-state HHG.

**Conclusion.** In brief, we propose a solid-state HHG amplification framework that encompasses the construction of an on-site gain/loss balanced diatomic lattice using a four-level system, the employment of k-gap engineering to demonstrate amplified Bloch oscillations, and the development of quadrupole radiation to calculate the amplified harmonic generations. Our approach specifically targets the amplification of EUV HHG in solids through k-gapped band structure engineering. The interband radiation modulated by k-gaps, significantly enhances the efficiency and spectral characteristics of the emissions. Despite the experimental challenges, our framework introduces a novel concept in k-gap engineering that opens new avenues for advanced solid-state photonics applications and the further exploration of compact, coherent EUV sources.

**Acknowledgement**

Y.P. acknowledges the support of the NSFC (No. 2023X0201-417-03) and the fund of the ShanghaiTech University (Start-up funding). The authors declare no competing financial interests.

Correspondence and requests for materials should be addressed to Y.P.
yiming.pan@shanghaitech.edu.cn